\documentclass[twocolumn]{aastex631}

\usepackage{comment}

\newcommand{\co}{${\rm ^{12}CO}\ J=3-2\ $}
\newcommand{\kms}{{\rm km\ s^{-1}}}
\newcommand{\mJB}{{\rm mJy\ {beam}^{-1}}}
\newcommand{\gcmt}{{\rm g\ {cm}^{-2}}}
\received{August 13, 2022}
\revised{September 5, 2022}
\accepted{September 6, 2022}

\submitjournal{ApJL}

\begin{document}

\title{Discovery of Line Pressure Broadening and Direct Constraint on Gas Surface Density in a Protoplanetary Disk}

\shorttitle{Discovery of Line Pressure Broadening in a Protoplanetary Disk} 
\shortauthors{Yoshida et al.}

\correspondingauthor{Tomohiro C. Yoshida}
\email{tomohiroyoshida.astro@gmail.com}
\author[0000-0001-8002-8473]{Tomohiro C. Yoshida}
\affiliation{National Astronomical Observatory of Japan, 2-21-1 Osawa, Mitaka, Tokyo 181-8588, Japan}
\affiliation{Department of Astronomical Science, The Graduate University for Advanced Studies, SOKENDAI, 2-21-1 Osawa, Mitaka, Tokyo 181-8588, Japan}

\author[0000-0002-7058-7682]{Hideko Nomura}
\affiliation{National Astronomical Observatory of Japan, 2-21-1 Osawa, Mitaka, Tokyo 181-8588, Japan}
\affiliation{Department of Astronomical Science, The Graduate University for Advanced Studies, SOKENDAI, 2-21-1 Osawa, Mitaka, Tokyo 181-8588, Japan}

\author[0000-0002-6034-2892]{Takashi Tsukagoshi}
\affiliation{Faculty of Engineering, Ashikaga University, Ohmae-cho 268-1, Ashikaga, Tochigi, 326-8558, Japan}

\author[0000-0002-2026-8157]{Kenji Furuya}
\affiliation{National Astronomical Observatory of Japan, 2-21-1 Osawa, Mitaka, Tokyo 181-8588, Japan}

\author[0000-0003-4902-222X]{Takahiro Ueda}
\affiliation{Max-Planck Institute for Astronomy (MPIA), K\"{o}nigstuhl 17, D-69117 Heidelberg, Germany}
\affiliation{National Astronomical Observatory of Japan, 2-21-1 Osawa, Mitaka, Tokyo 181-8588, Japan}



\begin{abstract}
The gas surface density profile of protoplanetary disks is one of the most fundamental physical properties to understanding planet formation. 
However, it is challenging to determine the surface dunsity profile observationally, because the H$_2$ emission cannot be observed in low-temperature regions.
We analyzed the Atacama Large Millimeter/submillimeter Array (ALMA) archival data of the \co line toward the protoplanetary disk around TW Hya, and discovered extremely broad line wings due to the pressure broadening.
In conjunction with a previously reported optically thin CO isotopologue line, the pressure broadened line wings enabled us to directly determine the midplane gas density for the first time.
The gas surface density at $\sim5$ au from the { central star} reaches $\sim 10^3\ {\rm g\ cm^{-2}}$, which suggests that the { inner region of the} disk has enough mass to form a Jupiter-mass planet.
{ Additionally, the gas surface density drops at the inner cavity by $\sim2$ orders of magnitude compared to outside the cavity.}
We also found a low CO abundance of $\sim 10^{-6}$ with respect to H$_2$, even inside the CO snowline, which suggests conversion of CO to less volatile species.
Combining { our results with} previous studies, the gas surface density jumps at $r\sim 20$ au, suggesting that the inner region ($3<r<20$ au) might be the magnetorotational instability dead zone.
This study sheds light on direct gas-surface-density constraint without assuming the CO/H$_2$ ratio using ALMA.
\end{abstract}

\keywords{Protoplanetary disks (1300) --- Astrochemistry (75)}


\section{Introduction} \label{sec:intro}
Gas mass and surface density of protoplanetary disks are crucial parameters for understanding planet formation processes.
Nevertheless, it is challenging to properly measure them.
Since the most abundant molecule, $\rm H_2$, does not radiate efficiently at low temperatures, other tracers need to be employed.
Traditionally, dust continuum emission has been used as in the case of the molecular clouds. 
However, the conversion of dust mass to gas mass is not straight forward, since the dust-to-gas mass ratio { is expected to change over the lifetime of the disk.}
CO isotopologues have been also employed, but recent observations suggest that CO is chemically depleted in some disks, and not a reliable tracer \citep[e.g.,][]{miot17}.
A combination of CO isotopologues and $\rm N_2H^+$ would be a candidate but detailed chemical modeling is needed \citep{trap22}.
So far, the most promising method of measuring gas mass and surface density profile is to use CO isotopologues and HD \citep[e.g.,][]{berg13, schw16, zhan17}.
However, the strong temperature dependency of the HD transitions makes the conversion to the mass uncertain.
Futhermore, HD emission lines at far-infrared are inaccessible by current instruments.

The protoplanetary disk around TW Hya is the nearest \citep[$D\sim 60.14\pm0.05$ pc;][2022, in prep]{gaia16}, and one of the most well-studied disks.
HD lines were detected toward the disk, but the estimated masses range over more than one order of magnitude \citep[][references are therein]{miot22}.

In this Letter, we report detection of the pressure broadened line wing in the \co spectra of the TW Hya disk.
The pressure broadening must be owing to highly dense gas near the midplane in the inner region, which allowed us to directly constrain the midplane gas density for the first time.
We introduce archival observations and show results and describe a parameterized model fitting in Section \ref{sec:res}.
We discuss the results in Section \ref{sec:dis} and summarize the study in Section \ref{sec:sum}.

\section{Observations and Results} \label{sec:res}

\subsection{Observations}
We obtained three data sets of the \co line in the TW Hya disk from the Atacama Large Millimeter/submillimeter Array (ALMA) science archive {(Project IDs: 2015.1.00686.S, 2016.1.00629.S, and 2018.1.00980.S )}.
We describe the data reduction details in Appendix \ref{app:obs}.
From the self-calibrated visibilities, we generated an imagecube with beamsize of $0\farcs 077 \times 0\farcs 058$ and averaged intensity maps of red- and blue-shifted line wings.
The channel width of the imagecube is $0.25\ \kms$.
The velocity ranges to generate the averaged intensity maps are shown in Figure \ref{fig:spec}.
To CLEAN the averaged intensity maps, we produced two maps for each wing with adopting the Briggs weighting (robust$=0$) and the natural weighting.

We determined the disk geometrical parameters by fitting the geometrically-thin Keplerian rotation model to a centroid velocity map with the \texttt{eddy} package \citep{eddy}.
The centroid velocity map was generated from the datacube using the \texttt{bettermoments} package \citep{bm}.
The stellar mass $M_\star$, the systemic velocity $v_{\rm sys}$, and the position angle were constrained to be $0.84\ M_{\odot}$, $2.84\ \kms$, $152^\circ$, respectively, with fixing the disk inclination angle to $i=5.8^\circ$ by following \citet{teag19}.

\subsection{Pressure broadened line wings}
\begin{figure}[t]
    \epsscale{1.1}
    \plotone{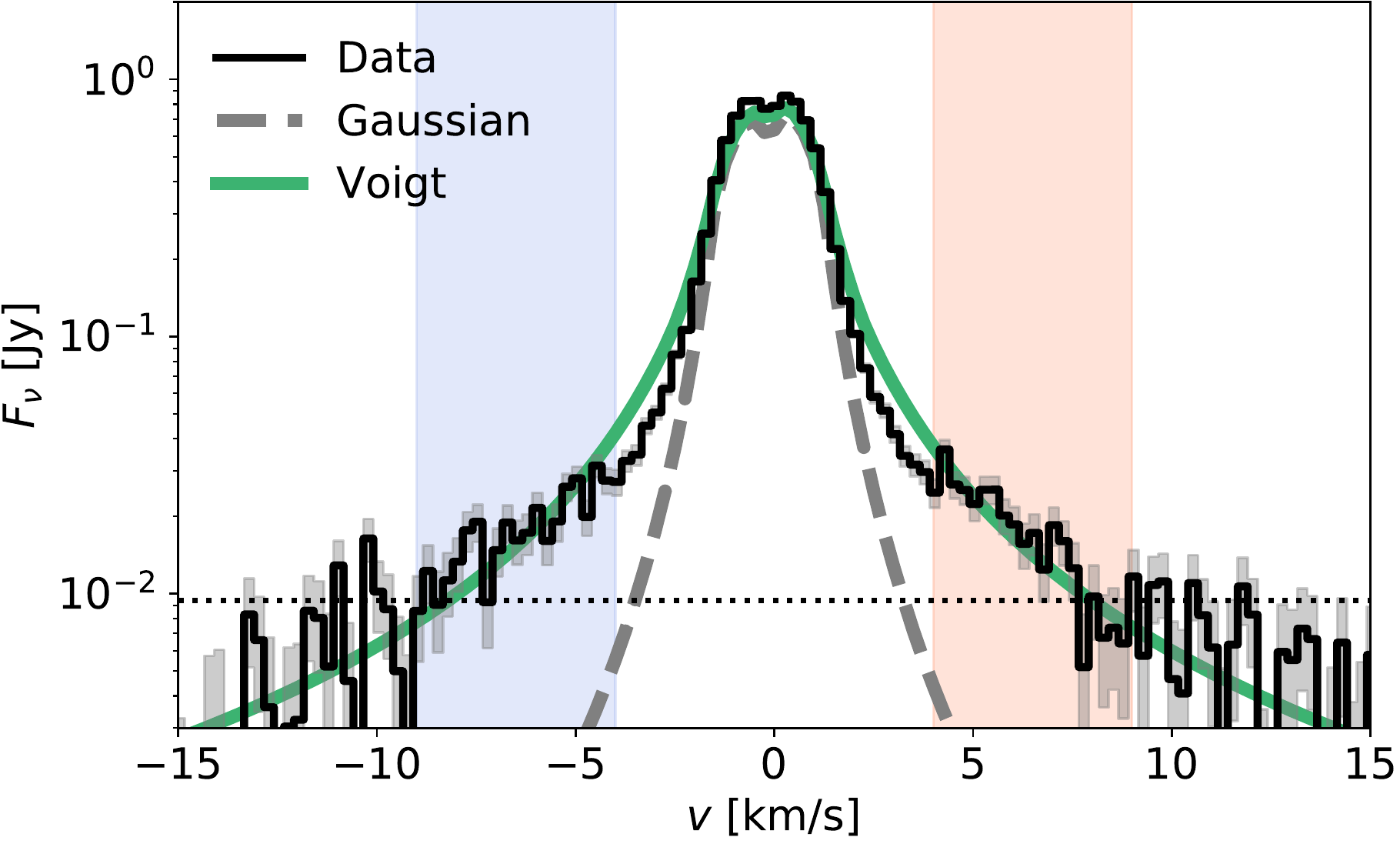}
    \caption{ Integrated spectrum of the \co line in the $0\farcs 2$ aperture (black solid line) with the uncertainty (gray shaded area) and a $3\sigma$ noise level (black dotted line).
    The gray dashed line indicates synthetic spectra of the \citet{huan18} model with a Gaussian line profile, while the green solid line shows the best-fit model results using the Voigt line profile (see Section \ref{sec:mod}).
    Blue- and red-masked ranges are used to create the averaged visibilities for each wing.
    $v$ denotes the velocity shift from the systemic velocity.
    \label{fig:spec}}
\end{figure}
Figure \ref{fig:spec} shows an integrated spectrum of the \co line with an aperture of $0\farcs 2$ in radius from the { star}.
The spectrum has broad line wings ranging $\pm10\ \kms$ from the systemic velocity ($2.84\ \kms$).
\citet{rose12} reported that the \co and $J=2-1$ lines exhibit emission up to $2.1\ \kms$ from the systemic velocity.
They were not able to reproduce this emission with simple disk models and proposed three possible explanations; a hot inner disk, a non-Keplerian velocity field, and a disk warp.
Since they used the earliest science verification data of ALMA, the sensitivity and spatial resolution were limited.
Therefore, the high-quality data obtained by combining the archival data{ , which has $\sim15$ times better point-source sensitivity,} enabled thorough investigation.

\begin{figure*}[ht!]
    \epsscale{1.2}
    \plotone{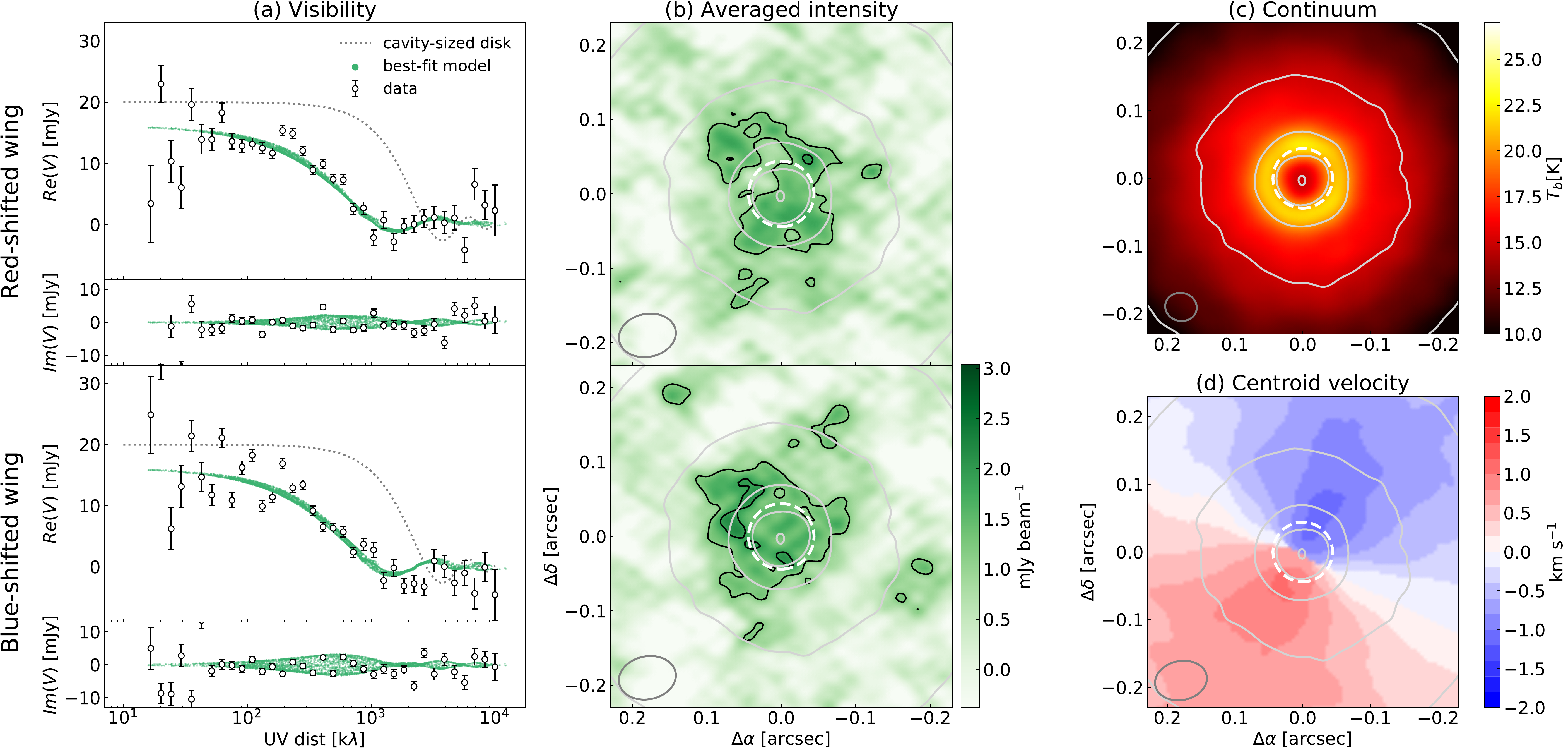}
    \caption{ 
    (a) uv-distance vs. azimuthally averaged complex visibilities (black circles). The green points and gray dotted lines indicate the best-fitted model and the cavity-sized uniform disk model, respectively. 
    (b) Averaged intensity maps with Briggs weighting (robust = 0) for the red- and blue-shifted line wings in the color scale and the black solid contour with 3 $\sigma$ and 5 $\sigma$ ($\sigma = 0.38\ \mJB$). The maps with natural weighting are shown in Figure \ref{fig:comparison}. 
    (c) 233 GHz continuum maps of \citet{tsuk19}. 
    (d) Centroid velocity map.
    The white dashed circles in panel (b), (c), and (d) indicate the inner cavity radius \citep{andr16}. The gray solid contour shows the dust continuum emission \citep{tsuk19}. The contour starts from the brightness temperature of 10 K with an interval of 5 K. 
    \label{fig:all}}
\end{figure*}
Firstly, we can exclude the possibility of the hot inner disk.
To broaden the line width to $\sim5\ \kms$, the gas temperature should be as high as $\sim10^5$ K.
However, such a high temperature thermally dissociates CO molecules.
In Figure \ref{fig:all} { (a)}, we plotted the azimuthally averaged visibility data with a synthetic visibility of an uniform disk with the cavity radius.
It is evident that the CO { line wing} emission is more extended than the cavity.
Figure \ref{fig:all} { (b)} shows the averaged intensity maps at the red- and blue-shifted wings ($4 < |v| < 9\ \kms$; see Figure \ref{fig:spec} ).
It was found that the { emission with $>3\sigma$ detection} at both line wings is extended up to $\sim0\farcs 1$ from the { star}, which is larger than the submillimeter innermost cavity { radius of $\sim0\farcs 04$} \citep{andr16}.
{ For reference, the 233 GHz continuum image of the same region is shown in Figure \ref{fig:all}(c).}
Note that we also show the averaged intensity maps with natural weighting in Figure \ref{fig:comparison}.
{ Notably, the emitting regions of both blue- and red-shifted components are spatially located around the star.
This means that the wing components behave differently from the Keplerian rotation with a narrow intrinsic line width for which the blue- and red-shifted components are expected to emit off-centered along the major axis of the disk.}
In addition, in the centroid velocity map { generated from the observed datacube} (Figure \ref{fig:all}{ d}), we can see that the centroid velocity shows a Keplerian velocity distribution, and does not exceed a few $\kms$ from the systemic velocity even in the position where the wing emission is detected. 
These results indicate that the line wing is not because of either the inner disk geometry or disk kinematics.

In the TW Hya disk, a { blue-shifted} photo-evaporation wind has been detected in the {[Ne II]} line \citep{pasc11}.
The spatially extended high-velocity CO emission could be a { molecular counterpart} of the wind.
However, at this frequency, highly optically thick dust disk lies in the midplane from the inner cavity edge to $\sim20$ au \citep{ueda20, maci21}.
Therefore, even if the CO disk wind exists, the red-shifted component beyond the cavity will not be observed.

The emission should be optically thin, because the brightness temperature is only $\sim3$ K, while the CO freezing temperature is $\sim17-27$ K \citep{qi13, zhan17}, and the midplane temperature of the continuum disk at $r \sim 13-41$ au is estimated to be similar \citep{ueda20, maci21}.

Instead of the above hypotheses to explain the broad line wings, we propose pressure broadening.
Pressure broadening can be ignored in most regions of protoplanetary disks, because the gas number density is too low to induce it.
However, in the inner region of disks, the gas density can be substantially high, and therefore, the pressure broadened wings become observable, especially in highly optically thick lines (at the line center).
Since the line wings should be optically thin and originate from well inside the CO snow line \citep{zhan17}, they directly trace dense gas near the midplane, and we can derive the gas surface density from the \co line and an optically thin CO isotopologue line without assuming the CO/H$_2$ ratio.

\subsection{Parameterized model fitting} \label{sec:mod}

To derive the gas surface density distribution, we constructed a simple slab model, and compared it with the observational data.
We assume a homogeneous isothermal slab along the line of sight at each radius in an axisymmetric disk.
These are reasonable assumptions, because the optically thin emission should trace near the disk midplane, the disk is nearly face-on, and no significant non-axisymmetric structure is observed in this region of both the line and continuum disks.

The specific intensity of the \co line at each radius at a velocity shift $v$ is given as
\begin{equation}
    I(v) = f B(T) ( 1-e^{-\tau(v)} ).
\end{equation}
Here, $B(T)$ and $\tau(v)$ denote the Planck function at temperature $T$ and the optical depth at each radius and $v$, respectively. 
We used the temperature profile proposed by \citet{ueda20}, $T = 30\ (r/10\ {\rm au})^{-0.4}$ K { for all radii}, which is consistent with that derived from multiwavelength dust continuum observations by \citet{maci21}.
The continuum emission at this frequency is optically thick at $r_{\rm cav, d}<r<20$ au, where $r_{\rm cav, d}$ is the inner cavity radius of the continuum disk, $2.7$ au \citep{andr16, maci21}.
$f$ is introduced to express the optical depth effect of the continuum disk at the midplane \citep{bosm21}, that is,
\begin{eqnarray} \label{eq:scat}
    f = \left\{
    \begin{array}{ll}
    1 &\  (r\leq r_{\rm cav, d}) \\
    1-\chi_d &\ (r>r_{\rm cav, d}).
    \end{array}
\right.
\end{eqnarray}
Here, $\chi_d$ is the intensity reduction coefficient given as $\sqrt{1-\omega}$ when the single scattering albedo $\omega$ is large \citep{zhu19}.
We followed \citet{ueda20} and adopted $\omega=0.88$.
The line optical depth $\tau(v)$ at $r>r_{\rm cav, d}$ is expressed as
\begin{equation}
    \tau(v) = g \tau_0 \phi(v-v_s),
\end{equation}
where $\tau_0$ is the optical depth of \co at the line center at each radius, and $\phi$ is the Voigt line profile normalized by the peak value.
The Voigt profile has two characteristic line widths; the Doppler width $\Delta v_D$ and the pressure width $\Gamma$, which is proportional to the H$_2$ number density at the midplane, $n_{\rm H_2}$, and $v_s$ is the line-of-sight velocity of the line center. 
The detailed formulation is described in Appendix \ref{appendix}.
We assumed the Keplerian rotation given by
\begin{equation}
    v_s = \sqrt{\frac{G M_\star}{r}} \cos{\theta} \sin{i},
\end{equation}
with $G$ and $\theta$ being the gravity constant and the azimuthal angle from the disk major axis, respectively.
The factor $g$ accounts for the gas depletion in the inner cavity;
\begin{equation}
    g = \left\{
    \begin{array}{ll}
    \delta_{\rm cav} &\  (r\leq r_{\rm cav, g}) \\
    1 &\ (r>r_{\rm cav, g}),
    \end{array}
    \right.
\end{equation}
where $r_{\rm cav, g}$ denotes the gas inner cavity radius.
At $r\leq r_{\rm cav, d}$, the { line} optical depth is doubled, since the continuum disk becomes optically thin in contrast to $r > r_{\rm cav, d}$.

To derive $\tau_0$, we first re-derived the ${\rm ^{13}C^{18}O}\ J=3-2$ line optical depth, using the data in \citet{zhan17} and considering the effect of dust scattering and the temperature profile of \citet{ueda20}.
Here, we adopted the line transition parameters from the LAMDA \citep{lamda} and HITRAN databases \citep{hitran} for the $\rm ^{12}C^{18}O$ line and the $\rm ^{13}C^{18}O$ line, respectively.
Then, the optical depth of $\rm ^{13}C^{18}O$ was converted to that of $\rm ^{12}C^{16}O$, assuming { isotope ratios of ${\rm ^{12}C/^{13}C=69}$ and ${\rm ^{16}O/^{18}O=557}$} \citep{wils99}.
The $\rm ^{13}C^{18}O$ optical depth at the line center becomes $\sim 0.6$ at $r<20$ au.
Although \citet{zhan17} constrained the optical depth profile only in $5-21$ au, we simply extrapolate it to the inner radius. 

The gas surface density at each radius is given as
\begin{equation}
    \Sigma_g = g \Sigma_{g, {\rm cav}} \left( \frac{r}{r_{\rm cav, g}} \right)^{-\gamma},
\end{equation}
where $\Sigma_{g, {\rm cav}}$ is the gas surface density at the gas cavity radius $r_{\rm cav, g}$.
Assuming vertical hydrostatic equilibrium, $n_{\rm H_2}$ is expressed as
\begin{equation}
    n_{\rm H_2} = \frac{\Sigma_g}{\mu m_p \sqrt{2 \pi } H_g}.
\end{equation}
Here, $m_p$ is the proton mass, and $H_g$ denotes the gas scale height
\begin{equation}
    H_g =  \frac{c_s}{\Omega_k},
\end{equation}
where $c_s$ and $\Omega_k$ are the sound speed and the Keplerian frequency, respectively.

\begin{figure*}[t]
    \epsscale{1.0}
    \gridline{\fig{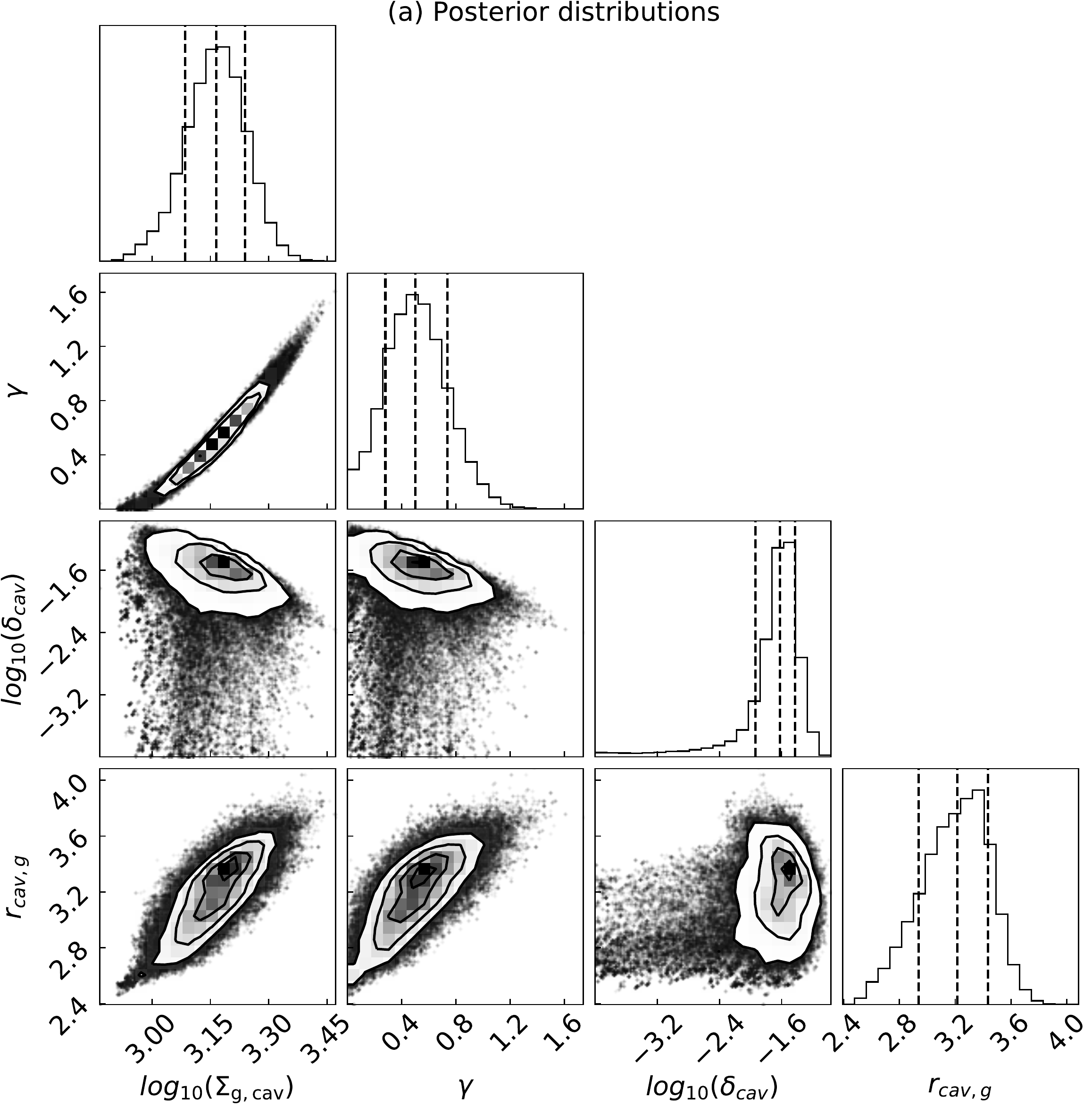}{0.52\textwidth}{}
            \fig{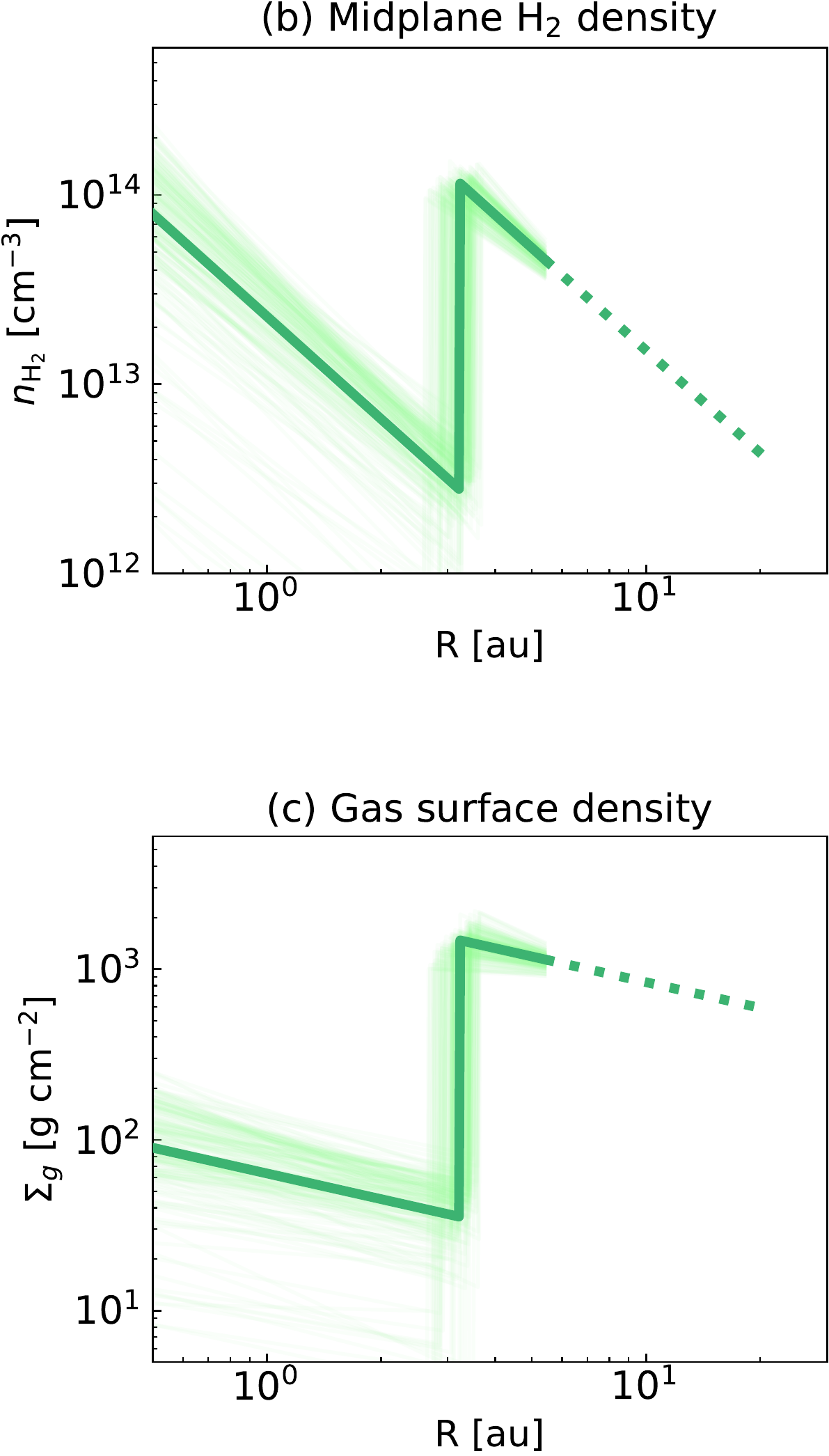}{0.3\textwidth}{}}

    \caption{(a) Posterior distributions of the model parameters. (b) The best-fit midplane $\rm H_2$ density. (c) The best-fit gas surface density. The green solid lines show the best-fit model. The models with parameters randomly selected from the posterior distributions are plotted in light green. 
    \label{fig:mcmc}}
\end{figure*}

Our specific intensity model has four free parameters; $\Sigma_{g, {\rm cav}}$, $\gamma$, $\delta_{\rm cav}$, and $r_{\rm cav, g}$.
After the model was generated and projected to the plane of the sky, we spectrally averaged the specific intensity distribution in $4<|v|<9\ \kms$ for the red- and blue-shifted wings.
The averaged intensity maps were sampled by the observational $(u,v)$ points in the Fourier domain using the \texttt{GALARIO} library \citep{galario}.
Then, we directly compared the models with observed visibilities.
We used the Markov Chain Monte Carlo (MCMC) method to sample the posterior probability distribution for each parameter.
Practically, we used the \texttt{emcee} package \citep{emcee} with chains of 256 walkers and 1024 steps.
We selected $-0.5(\chi_{\rm red}^2 + \chi_{\rm blue}^2)$ as the log likelihood function, where $\chi^2$ denotes the chi-squared between the visibility model and the observed visibility, and the subscripts indicate each line wing.
The best-fit parameters with uncertainties are $\log_{10}(\Sigma_{g, {\rm cav}}) = 3.17^{+0.07}_{-0.08}\ \gcmt$, $\gamma = 0.50^{+0.24}_{-0.22}$, $\log_{10}(\delta_{\rm cav}) = -1.62^{+0.19}_{-0.32}$, and $r_{\rm cav, g}=3.21^{+0.22}_{-0.28}$ au.
The posterior distributions, the best-fit midplane $\rm H_2$ density profile, and the best-fit gas surface density profile are shown in Figure \ref{fig:mcmc}.

\begin{figure*}[t]
\epsscale{1.0}
\plotone{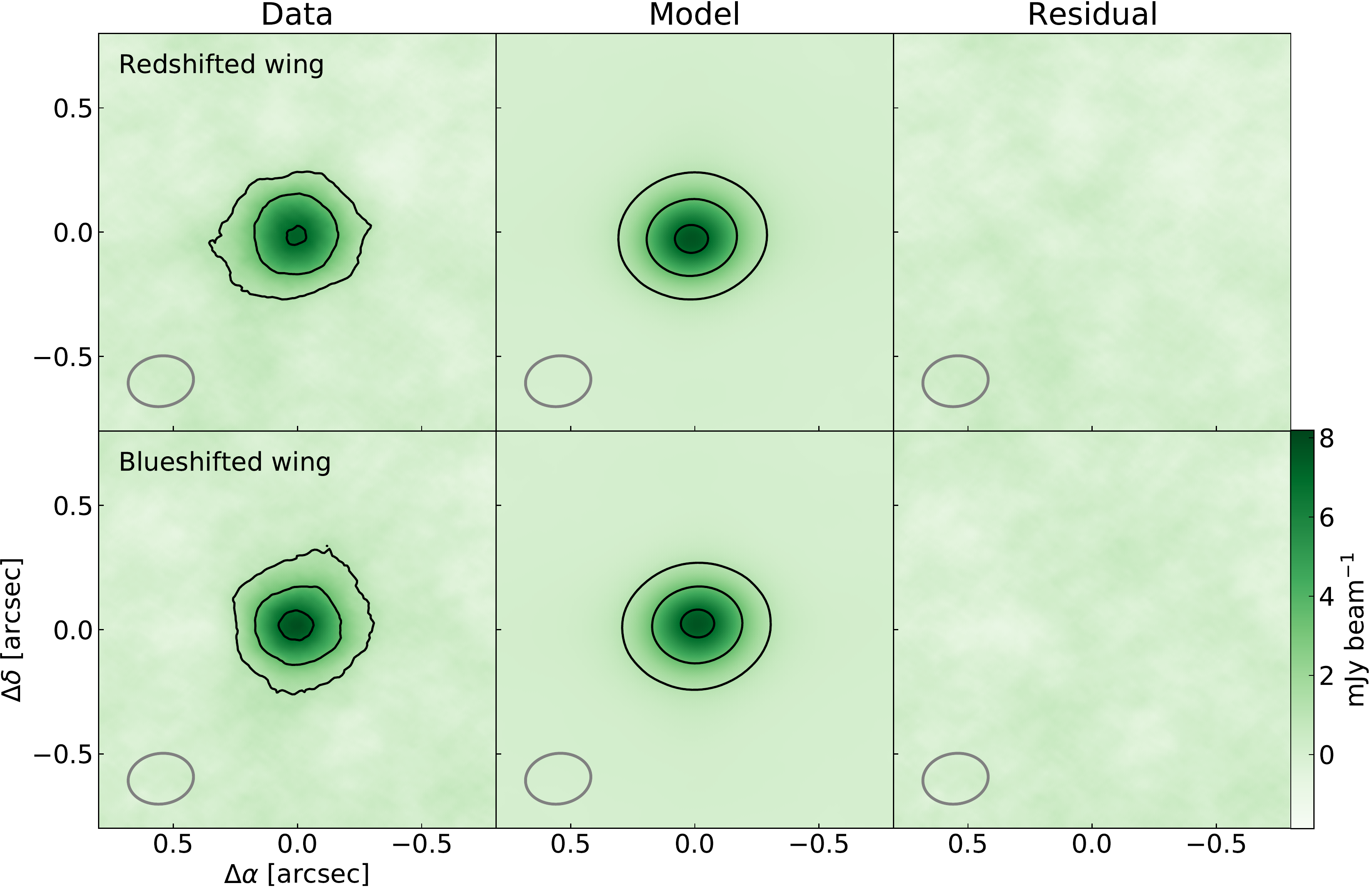}
\caption{Comparison of the observed averaged intensity maps with natural weighting and best-fit models. The residual maps are also plotted. The black contours start from $5\sigma$ with an interval of $5\sigma$ with $\sigma=0.23\ \mJB$.
We also show maps with finer resolution in Figure \ref{fig:all}(b).
\label{fig:comparison}}
\end{figure*}

Figure \ref{fig:comparison} shows the averaged intensity maps with the natural weighting, model images, and residual images calculated in the visibility domain and Fourier-transformed.
Additionally, we plot the complex visibilities against the uv-distance in green points in Figure \ref{fig:all}(b).
The model is well fitted to the observations, and the free parameters are well constrained.

\begin{figure*}[t]
    \epsscale{1.1}
    \plotone{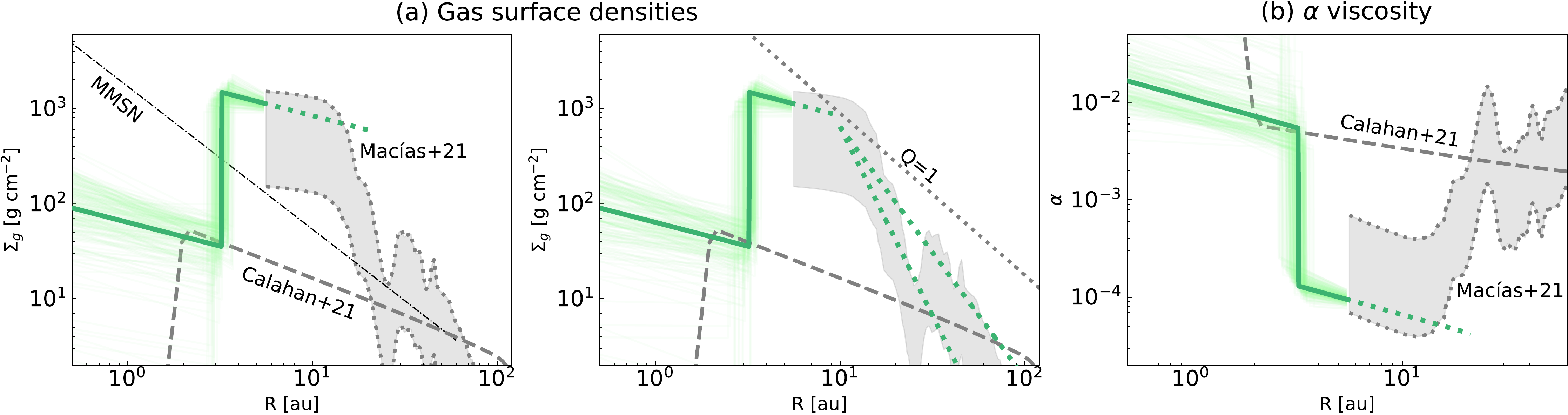}
    \caption{{ (a) Radial profiles of the gas surface density. The green solid lines show the best-fit model. The models with parameters randomly selected from the posterior distributions are plotted in light green. The MMSN model and the \citet{cala21} model are also shown in gray lines.
    The gray dotted lines indicate the gas surface density converted from the dust surface density of \citet{maci21}, assuming a gas-to-dust surface density ratio of 10 and 100. In the right panel, we also show the gas surface density when the Toomre Q equals unity, and the gas surface density profiles starting from $r=10$ au of the best-fit profile with fixing the total gas masses to 0.023 $M_\odot$ or 0.06 $M_\odot$. 
    (b) The $\alpha$ viscosity parameter for each gas surface density profile, assuming a steady accretion disk. }
    \label{fig:radprof}}
\end{figure*}
Based on the averaged intensity model, we found that half the total flux of each wing originates from $r < 6$ au.
Therefore, we assume that the derived radial profile is robustly constrained only for $r \lesssim 6 $ au.
We also note that the line wing emission would not be sensitive to $r \sim 1-3$ au, because the midplane density might be significantly lower than those in both the innermost region and outside of the cavity (Figure \ref{fig:radprof}b).
Even though, our model is well fitted under the assumption of the single { power law} index { over the fitted region ($r<5$ au)}.

We also generated synthetic image cubes from the model proposed by \citet{huan18} with the Gaussian line profile, and a 2D model based on the best-fit parameters with the Voigt line profile using RADMC-3D \citep{dull12}.
To generate the 2D model, the midplane temperature of the \citet{huan18} model was replaced with our assumed values, and the CO surface density, which is based on \citet{zhan17}, was increased by a factor of $1.3$ to cancel the modification on the midplane temperature for maintaining the same intensity.
The integrated spectra are plotted in Figure \ref{fig:spec}.
While both models are matched to the line core, only the Voigt profile model can reproduce the line wings.
Note that our model is not fine-tuned and the continuum emission is not taken into account, which might lead to errors, especially at $v\sim4\ \kms$.

We rely on the extrapolation of the $\rm ^{13}C^{18}O$ optical depth profile derived by \cite{zhan17} to $r<5$ au.
The intensity at the line wings are proportional to $\tau_0 n_{\rm H_2} \propto X_{\rm CO} n_{\rm H_2}^2$ (Appendix \ref{appendix}), with $X_{\rm CO}$ being the CO abundance.
{ Therefore, if $\tau_0$ is higher than the assumed values, then $n_{\rm H_2}$ would need to be smaller to still reproduce the intensity in the line wings, which would also imply a higher CO abundance.
Supposing that $n_{\rm H_2}$ were $\sim 30$ times smaller than the best-fit profile in reality (i.e., similar to the interpolation of \citet{cala21}, see Section \ref{sec:sd}), $X_{\rm CO}$ would be $\sim 10^{3}$ times larger than the best-fit results ($\sim 6.5\times 10^{-7}$; see Sec.\ref{sec:sd}) at $r< 5$ au.}
However, it is unrealistic that the CO abundance has such an extreme jump much inside the CO snowline, and becomes even six times larger than the interstellar medium value.
Moreover, since the carbon-poor gas is suggested in both the inner cavity \citep{bosm19, mccl20} and the outer disk \citep{zhan17, cala21}, it is unlikely that the CO/H$_2$ has significantly different values only at $r\sim 5$ au.
Therefore, the existence of the dense gas at $r\sim$ 3--5 au is robust.
However, higher special resolution observations of optically thin CO isotopologue lines will enable more accurate estimates.

{ The optically thin emission of the line wings may originate from the dust $\tau=1$ surface. Therefore, our results could be underestimated since the gas below the dust $\tau=1$ surface is unobservable. However, as long as the $\tau=1$ surface of the dust continuum emission at 346 GHz locates under the gas scale height, this effect will be less than a factor of two.}

{ Inside the cavity, the midplane gas temperature could be higher than the extrapolation from the dust temperature outside of the cavity since the gas and dust temperatures can be decoupled \citep[e.g.,][]{brud13}. 
To examine this effects, we also run the MCMC fitting with enhancing the temperature inside cavity by a factor of three. We find consistent values as before except that $\delta_{\rm cav}$ has a larger uncertainty.
}

\section{Discussion} \label{sec:dis}
\subsection{The gas surface density profile} \label{sec:sd}

The resulting radial profile of the gas surface density is compared with other models in Figure \ref{fig:radprof}(a).
The gas surface density at $r = 5$ au is $\sim1.2\times10^{3}\ \gcmt$, which is $\sim8$ times larger than the minimum mass solar nebula \citep[MMSN;][]{haya81} at the same location.
On the basis of the surface density profile, we found that the gas mass inside $\sim5$ au is $\sim7\ M_J$.
Recently, dust continuum modeling including scattering effects suggests that the dust surface density at $r=5$ au reaches $\sim14\ \gcmt$ \citep{ueda20, maci21}, which is also $\sim 5$ times larger than that from the MMSN.
Therefore, we conclude that the inner region ($r<5$ au) of the TW Hya disk still has the ability to form a Jupiter-mass planet in terms of material amount, although the central star is relatively old \citep[3$-$10 Myr; ][]{barr06, vacc11}.

We calculated the Toomre Q parameter \citep{toom64} according to
\begin{equation}
    Q = \frac{c_s \Omega_k}{\pi G \Sigma_g}.
\end{equation}
The Toomre Q parameter at $r=5$ au is $\sim2.4$.
Therefore, the disk should be gravitationally stable at least up to $r<5$ au, which is consistent in the absence of spiral arms excited by the gravitational instability.
However, if we simply extrapolate the power law beyond $r=5$ au, the Toomre Q value reaches 1 at $\sim10$ au.
This implies that the gas surface density beyond 5 au should be lower than the extrapolation, or the cooling time scale is long enough to suppress the gravitational instability due to the high optical depth \citep{armi10}.

By comparing the gas surface density with the CO surface density { derived from the $\rm ^{13}C^{18}O$ line optical depth}, we found the gas-phase $\rm CO/H_2$ ratio $X_{\rm CO}$ of $\sim 6.5\times10^{-7}$ at $r \sim 5$ au, which is $\sim150$ times lower than the interstellar medium value.
It is known that the gas-phase CO is strongly depleted by $1-2$ orders of magnitude in the outer region \citep[e.g.,][]{zhan19, cala21}, as well as in the inner cavity \citep{bosm19, mccl20}.
Our results show that CO depletion may be extreme even inside the CO snowline outside the cavity.
This is in line with a scenario proposed by \citet{bosm19}, where CO is converted to less volatile species, such as $\rm CO_2$ and $\rm CH_3OH$, locked to the large dust grains, and trapped to the innermost submillimeter ring \citep{andr16}.
Interestingly, \citet{ueda20} suggested that the inner region of dust disks are dominated by small $\sim 300\ {\rm \mu m}$ grains and attributed it to the poor stickiness of icy grains covered by $\rm CO_2$ ice, as shown in experiments \citep{musi16}.

Using the resulting gas surface density profile, we can estimate the inner disk ($r < r_{\rm cav, g}$) gas mass of $\sim 1.7 \times10^{-4}\ M_\odot$, which is in excellent agreement with previous estimates by modeling infrared lines including $\rm H_2$ \citep{bosm19}.

\subsection{Gas surface density jump at $r\sim 20$ au} \label{sec:coabn}

The best-fit surface density at $r=5$ au is $\sim30-40$ times larger than previous estimates on the surface density profile \citep{zhan17, cala21}.
These surface density profiles are based on the spatially resolved CO isotopologue lines and the spatially unresolved HD $J=1-0$ line.
\citet{trap17} found that an emitting region of the HD $J=1-0$ line ranges $9-70$ au in radius, indicating that the detected flux mostly arises from the relatively outer region, although \citet{zhan17} considered that 90 \% of the flux originates from the inner 20 au.
As the CO line wing emission has sensitivity at $r \lesssim 5$ au, our results would be able to constrain the gas surface density profile independently and may not be inconsistent with previous estimates which depend on the HD emission from the outer regions.

Comparing our results with the dust surface density profile derived by \citet{ueda20}, the gas-to-dust mass ratio at $r\sim5$ au can be calculated to be $\sim84$.
\citet{maci21} derived the dust surface density profile at $r > 5$ au, which jumps at $r\sim$ 20 au.
The profile inner $r \sim 10$ au is consistent with \citet{ueda20}.
The gas-to-dust ratio at $r>20$ au can be estimated to be $30-70$ from the \citet{maci21} dust surface density profile, considering the uncertainty in disk gas mass \citep[$0.02-0.06\ M_\odot$;][]{berg13, kama16, trap17}.
{ The results that the gas-to-dust ratio at $\sim 5$ au ($\sim 84$) is consistent with that at $r>20$ au ($30-70$) within a factor of three imply that the gas surface density exhibits a two-order-of-magnitude jump between $r \sim 5$ au and $> 20$ au similar to the dust surface density.}

{ We can also constrain the gas surface density profile at $\sim 5-20$ au in terms of dynamical stability. 
In Figure 5(a, right), we plot the gas surface density with which the disk is gravitationally unstable ($Q=1$). 
This would provide an upper limit of the gas surface density.
The actual profile should be below the Q=1 curve, which is consistent with the gas surface density obtained by \citet{maci21} with assuming the gas-to-dust ratio of $10-100$. 
We note that the gas-to-dust ratio may vary locally and the gas density jump at $\sim20$ au might be less pronounced.
However, if gas-to-dust ratio is much higher than 100 at $r \gtrsim 20$ au, the gas disk mass would be inconsistent with previous observations \citep{berg13, trap17} as shown bellow.
}

{ We calculated a power law gas surface density profile starting from $r=10$ au, where the extrapolation of the best-fit profile reaches $Q=1$, with fixing the mass within $10<r<200$ au to $0.023\ M_\odot$ \citep{trap17} and $0.06\ M_\odot$ \citep{berg13}. 
These masses are lower and upper side values estimated from the HD observations which are sensitive to the outer regions rather than $r<10$ au. 
The derived profiles are shown in the right panel of Figure \ref{fig:radprof}(a), which overlaps the profile based on dust with the gas-to-dust ratio of $10-100$.
However, we note that the power-law profiles at $r \gtrsim 30$ au are not real because they are inconsistent with the previous analysis of HD and CO \citep{cala21}.
}

Also, it is suggested that the gas-to-dust ratio could be moderately modified from the interstellar value ($\sim 100$) over the disk.
This implies moderate gas-depletion and/or that the continuum gaps \citep[e.g.,][]{tsuk16} did not efficiently trap mm-dust grains.
When adopting a lower side of the gas-to-dust mass ratio at the outer region ($\sim 30$), the gas-to-dust mass ratio at $r \sim 5$ au would be higher than at $r>20$ au by a factor of three.
These gas-to-dust ratios means that the gas is more depleted in the outer region than the inner region. Otherwise, the large grains are depleted in the inner region than in the outer region.
Since the gas depletion mechanism such as the disk wind should be more effective in the inner region and the radial drift of dust grains decreases the ratio in the outer region, the former may be not preferable.
However, as suggested by the CO depletion factor, the inner region ($3<r<20$ au) should be a highly efficient dust trap \citep{bosm19}.
The high gas-to-dust ratio { at $r \sim 5$ au}, therefore, might imply that there are substantial hidden masses or planetesimals in this region.

\subsection{Implications for the MRI dead zone}
It is believed that the magnetorotational instability (MRI) is suppressed at relatively inner regions of disks, which are called as dead zones \citep[e.g.,][]{gamm96}.
The gas and dust density in the dead zone become higher than those in the active regions \citep{dzyu13, turn14}.
Therefore, the gas and dust surface density jump at $r \sim 20$ au can be interpreted as the dead zone.
We also calculated the $\alpha$ viscosity parameter (Figure \ref{fig:radprof}b) defined as
\begin{equation}
    \alpha = \frac{\dot{M}_{\rm acc} \Omega_k}{3 \pi c_s^2 \Sigma_g},
\end{equation}
assuming a steady-state accretion disk \citep{shak73} with a mass accretion rate of $\dot{M}_{\rm acc} = 1.5\times10^{-9}\ M_\odot {\rm yr^{-1}}$ \citep{bric12}.
$\alpha$ takes a value $\sim10^{-3}$ at $r> 20$ au, while $\sim 10^{-4}$ at $3< r <20$ au.
Recent MRI accretion models have shown that the dead-zone outer edge would be located at few tens of au with a steep transition of $\alpha$ from $\sim 10^{-4}$ to $\sim 10^{-3}$ \citep{dela22}, which is consistent with the $\Sigma_g$ and $\alpha$ profiles obtained in this work.

{ Another possibility to create the gas surface density jump is a gas giant planet. However, the significant asymmetry of dust surface density inside and outside of the dust gap at $\sim 26$ au is not observed in hydrodynamical simulations \citep[e.g.,][]{kana16}.
Also, the large amount of dust grains in $r<20$ au suggests the gap does not efficiently filter dust grains, implying that the large gas surface density jump is unlikely to originate from a giant planet. }

Our results show that the gas is depleted in the inner cavity, which enhances the $\alpha$ under a constant accretion rate.
This inner jump of the $\Sigma_g$ and $\alpha$ may be corresponded to the dead-zone inner edge.
However, the temperature should be as high as $\sim 1000$ K \citep{gamm96, desc15} to activate the MRI, which is unlikely for the TW Hya disk.
Alternatively, a planet can create the inner cavity by accreting disk material.
According to \citet{kana15}, the mass of the planet can be estimated from the disk aspect ratio, $\alpha$ viscosity parameter, and gap depth.
Using the best-fit model and assuming $\alpha=10^{-4}$, the possible planet mass at $r=2.5$ au is estimated to be $\sim9\ M_\earth$.
{ Additionally, the photoevaporation can be another process to open the cavity \citep{owen11, pasc11}. }

\section{Summary} \label{sec:sum}
We analyzed ALMA archival data of the \co line in the TW Hya disk.
Thanks to the high-sensitivity and high-spatial resolution, we detected broad line wings that extend over $\sim 20\ \kms$.
We attributed the broad line wings to the pressure broadening rather than the kinematics or temperature of the disk, because both red- and blue-shifted wing emissions are spatially extended and distributed around the { star}, and the central velocity map exhibits Keplerian rotation.
The pressure broadened line wings are direct evidence of dense gas near the midplane.
By fitting the simple parameterized model, we derived the gas surface density profile { in the inner region of the disk}.
{ On the basis of the best-fitted model, it is suggested that the gas surface density at $\sim5$ au from the star reaches $\sim 10^3\ {\rm g\ cm^{-2}}$. Thus, the inner region of the disk has enough mass to form a Jupiter-mass planet.}
Additionally, the CO/H$_2$ ratio is as low as $\sim 10^{-6}$ even inside the CO snowline, implying conversion of CO to less volatile species.
In addition, our results provide a new anchor point of the gas-to-dust mass ratio at $r \sim 5$ au.
In conjunction with the dust surface density profile, the gas surface density may jump at $r\sim 20$ au, which can be interpreted as the MRI dead zone.
Our results show that the pressure broadened line wings are capable of measuring gas mass and surface density of protoplanetary disks, which is complementary to other tracers sensitive to the outer regions such as the combination of CO isotopologues and N$_2$H$^+$.

\begin{acknowledgments}
We would like to acknowledge the anonymous referee for helpful remarks and comments.
This Letter makes use of the following ALMA data: ADS/JAO.ALMA\#2015.1.00686.S, 2016.1.00629.S, and 2018.1.00980.S. ALMA is a partnership of ESO (representing its member states), NSF (USA), and NINS (Japan), together with NRC (Canada), MOST and ASIAA (Taiwan), and KASI (Republic of Korea), in cooperation with the Republic of Chile. The Joint ALMA Observatory is operated by ESO, AUI/NRAO, and NAOJ.
T.C.Y. was supported by the ALMA Japan Research Grant of NAOJ ALMA Project, NAOJ-ALMA-265.
This work is supported by JSPS and MEXT Grants-in-Aid for Scientific Research, 18H05441, 19K03910, 20H00182 (H.N.), 20K04017 (T.T.), 20H05847 and 21K13967 (K.F.).
T.U. acknowledges the support of the DFG-Grant "Inside: inner regions of protoplanetary disks: simulations and observations" (FL 909/5-1).
\end{acknowledgments}

%

\vspace{5mm}
\facilities{ALMA}


\software{astropy \citep{astropy}, matplotlib \citep{plt}, GALARIO \citep{galario}, uvplot \citep{uvplot}, emcee \citep{emcee}, eddy \citep{eddy}, bettermoments \citep{bm}, CASA \citep{mcmu07} }



\appendix

\section{Data reduction}\label{app:obs}
We obtained observational data from the ALMA science archive.
The project IDs are 2015.1.00686.S (PI. S.Andrews), 2016.1.00629.S (PI. I. Cleeves), and 2018.1.00980.S (PI. R. Teague).
These observations are originally presented in \citet{andr16}, \citet{huan18}, and \citet{teag21}, respectively.
The uv-coverage is $\sim15-12300\ k\lambda$, and the total integration time reaches $\sim10.4$ hours.

The visibility data were calibrated using the provided scripts except for 2015.1.00686.S which was calibrated with the support of the ALMA East Asian Regional Center Helpdesk.
We performed the following data reductions using the Common Astronomical Software Application \citep[CASA;][]{mcmu07} package (modular version 6.4.3).
After generating the line-flagged visibilities, we employed six-round phase and one-round amplitude self-calibration iteratively using the CLEANed continuum image.
The solutions were applied to the \co line data.
The visibility was spectrally re-grided to a channel width of $0.25\ \kms$ using the CASA task \texttt{cvel2}, and the continuum emission was subtracted by fitting a linear function from $17.5$ to $25.0\ \kms$ and from $-12.5$ to $-20\ \kms$.
Then, the visibility was Fourier-transformed and CLEANed.
We adopted the auto mask implemented in CASA to mask the source regions, multi-scale deconvolution kernel of [0, 0.1, 0.25, 0.5, 1, 2, 4] arcsec, and Briggs weighting with a robust parameter of 0.
The final beam size is $0\farcs  077\times0\farcs  058$ with PA$ = -79^\circ$.
Finally, we applied the JvM correction \citep{jors95, czek21} to fairly evaluate faint emission.
The resulting RMS noise level is $\sim1.1\ \mJB$, which agrees with the previous publications.
We also generated averaged visibilities of line wing components for fitting models (see Section \ref{sec:mod}) and CLEANed them with the above parameters (robust $=0$) and natural weighting.
The averaged intensity images with natural weighting have a beam size of $0\farcs 26 \times 0\farcs 20$ with PA$=-82^\circ$ and noise level of $\sim0.23\ \mJB$.

\section{Formulation of the Voigt Line Profile}\label{appendix}
First, we define the Doppler width $\Delta v_D$ and pressure width $\Gamma$.
The Doppler width in the slandered deviation of the Gaussian component is written as
\begin{equation}
    \Delta v_D = \sqrt{\frac{k_B T}{m_{\rm CO}} + \Delta v_t^2},
\end{equation}
where $k_B$, $T$, $m_{\rm CO}$, and $\Delta v_t$ are the Boltzmann constant, gas temperature, CO molecular weight, and turbulent velocity.
We adopted $\Delta v_t = 0.01\ \kms$ because of weak turbulence at the outer disk \citep{teag16, flah18}, and this assumption would not affect the results, since we consider much higher velocity ranges.

The pressure width in the half width at half maximum is given as
\begin{equation}
    \Gamma = C_{\rm p}(T) n_{\rm H_2},
\end{equation}    
where $C_{\rm p}(T)$ and $n_{\rm H_2}$ are the temperature-dependent pressure broadening coefficient and the number density of ${\rm H_2}$ at the midplane, respectively.
We define $C_{\rm p}(T)$ as
\begin{equation}
    C_{\rm p}(T) \equiv \frac{c^2 k_B T}{\nu_0} \left\{ \zeta_{\rm H_2} \left( \frac{T_{\rm ref}}{T} \right)^{\xi_{\rm H_2}} + 0.19 \zeta_{\rm He} \left( \frac{T_{\rm ref}}{T} \right)^{\xi_{\rm He}} \right\},
\end{equation}
where $c$, $k_B$, and $\nu_0$ are the light speed, Boltzmann constant, and rest frequency of \co, respectively.
For the remaining parameters, we adopted line-shape parameters of the \co line broadened by collision with $\rm H_2$ and $\rm He$ as well as $T_{\rm ref} = 293$ K from the HITRAN database. Note that $\zeta$ and $\xi$ are given as $\gamma$ and $n$, respectively, in the HITRAN database.
We also assumed that a hydrogen to helium mass ratio of $\sim2.6$ and mean molecular weight per particle $\mu = 2.37$ \citep{kauf08}.
To calculate the Voigt profile, the python function \texttt{scipy.special.voigt\_profile} \citep{scipy} was used in practice.

The peak-normalized Voigt profile $\phi(v)$ can be expressed as
\begin{equation} \label{vp}
    \phi(v) = \frac{a}{\pi} \frac{e^{-a^2}}{{\rm erfc(a)}} \int_{-\infty}^{\infty} \frac{e^{-y^2}}{a^2+ (u-y)^2} dy,
\end{equation}
where
\begin{eqnarray}
    a &=& \frac{\Gamma}{\Delta {v_D}}, \\
    u &=& \frac{v}{\Delta {v_D}},
\end{eqnarray}
with $\rm erfc$ being the complementary error function \citep{rybi79}.
In the case of inner regions of protoplanetary disks, the pressure width $\Gamma$ is significantly smaller than the Doppler width $\Delta v_D$, and therefore, $a \ll 1$.
Additionally, $u \gg 1$ considering line wings.
Under physical conditions of protoplanetary disks and considering CO rotational transition lines, Eq.(\ref{vp}) asymptotically approaches
\begin{equation}
    \phi(v) \simeq \frac{a}{\sqrt{\pi}} \frac{e^{-a^2}}{{\rm erfc(a)}} u^{-2} \simeq \frac{a}{\sqrt{\pi} u^2}.
\end{equation}
Thus, the intensity of optically thin line wings from the homogeneous slab can be given as
\begin{equation}
    B(T)(1-e^{-\tau_0 \phi(v)}) \simeq B(T) \frac{ \tau_0 a}{\sqrt{\pi} u^2}.
\end{equation}
Since $\Gamma \propto n_{\rm H_2}$ and $\tau_0 \propto X_{\rm CO} n_{\rm H_2}$, the intensity is proportional to $\tau_0 n_{\rm H_2}$ or $X_{\rm CO} n_{\rm H_2}^2$.


\bibliography{sample631}{}
\bibliographystyle{aasjournal}



\end{document}